\documentclass[aps,twocolumn]{revtex4}

\usepackage{graphicx}
\usepackage{natbib}

\usepackage{graphics}
\usepackage{epsfig}
\usepackage{bm}
\usepackage{amsmath,amssymb}
\usepackage{stmaryrd}

\def\ov#1{\overline{#1}}

\def\vb#1{\mbox{\boldmath$#1$}}

\def\wh#1{\widehat{#1}}
\def\bdot{\,\vb{\cdot}\,}
\def\btimes{\,\vb{\times}\,}

\def\bhat{\wh{{\sf b}}}
\def\cal#1{\mathcal{#1}}

\def\bhat{\wh{{\sf b}}}

\newcommand{\bc}{\begin{center}}
\newcommand{\ec}{\end{center}}
\newcommand{\bt}{\begin{tabbing}}
\newcommand{\et}{\end{tabbing}}
\newcommand{\be}{\begin{equation}}
\newcommand{\ee}{\end{equation}}
\newcommand{\ba}{\begin{eqnarray}}
\newcommand{\ea}{\end{eqnarray}}

\begin{document}

\title{Comment on “Modification of Lie’s transform perturbation theory for charged particle motion in a magnetic field” [Phys. Plasmas, 30, 042515 (2023)]}

\author{A.J.~Brizard$^{a}$}
\affiliation{Department of Physics, Saint Michael's College, Colchester, VT 05439, USA \\ $^{a}$Author to whom correspondence should be addressed: abrizard@smcvt.edu}

\begin{abstract}
A recent paper by L.~Zheng [Phys. Plasmas, 30, 042515 (2023)] presented a critical analysis of standard Lie-transform perturbation theory and suggested that its application to the problem of charged-particle motion in a magnetic field suffered from ordering inconsistencies. In the present Comment, we suggest that this criticism is unjustified and that standard Lie-transform perturbation theory does not need to be modified in its application to guiding-center theory.
\end{abstract}

\date{\today}

\maketitle

In a recent paper, Zheng \cite{Zheng} suggested that when standard Lie-transform perturbation theory \cite{RGL_1982} is applied to the guiding-center theory of charged-particle motion in a magnetic field \cite{RGL_1981,RGL_1983,Brizard_1989}, ordering inconsistencies arise. Unfortunately, Zheng never defined an ordering parameter (denoted $\epsilon$ in the present Comment) in his critique of standard Lie-transform perturbation theory and, here, it is argued that his proposed modification of Lie-transform perturbation theory is completely unnecessary.

In guiding-center theory \cite{Northrop_1963,Cary_Brizard_2009}, the mathematical construction of the magnetic moment relies on the space-time scales $(L_{B},\omega^{-1})$ of the confining magnetic field ${\bf B} = B\,\bhat$ to be long compared to the characteristic gyroradius $\rho$ and the gyroperiod $\Omega^{-1} = mc/eB$, respectively, leading to the small dimensionless small parameter \cite{Cary_Brizard_2009}
\begin{equation}
\epsilon_{B} \;\equiv\; \rho/L_{B} \;\sim\; \omega/\Omega \;\ll\; 1,
\label{eq:ordering}
\end{equation}
which is also used in Zheng's paper. While this dimensional parameter makes physical sense, it is not an ordering parameter {\it per se} to be used in a perturbation expansion.

In early formulations of guiding-center theory \cite{Kruskal_1958,Northrop_1963,Kruskal_1965,Gardner_1966,Banos_1967}, the dimensional ratio $m/e$ was proposed as an ordering parameter in deriving guiding-center equations of motion, which is consistent with Eq.~\eqref{eq:ordering}, i.e., $\epsilon_{B} \propto m/e$. In previous Hamiltonian guiding-center models \cite{Brizard_1995,Cary_Brizard_2009,Tronko_Brizard_2015,Brizard_2023}, on the other hand, a dimensionless ordering parameter $\epsilon$ was introduced either as a mass renormalization $m \rightarrow \epsilon\,m$ \cite{Brizard_1995}, or as a charge renormalization $e \rightarrow e/\epsilon$ \cite{Cary_Brizard_2009,Tronko_Brizard_2015,Brizard_2023}, so that the dimensional ratio $m/e \rightarrow \epsilon\,m/e$ is indeed considered small in both cases (i.e., $\epsilon_{B} \sim \epsilon$). These renormalization orderings can then form the basis for a well-defined perturbation-expansion analysis of charged-particle motion in a magnetic field by Lie-transform perturbation methods \cite{RGL_1982}. 

Depending on the renormalization ordering used, we can begin our guiding-center perturbation analysis with the particle Lagrangian, either expressed according to the charge renormalization as
\begin{equation}
L({\bf x},{\bf p}) \;=\; \left(\frac{e}{\epsilon c}\,{\bf A} \;+\; {\bf p}\right)\bdot\dot{\bf x} \;-\; \left(\epsilon^{-1}\,e\,\Phi \;+\; \frac{|{\bf p}|^{2}}{2m}\right),
\label{eq:Lag_1}
\end{equation}
or, according to the mass renormalization, as
\begin{equation}
L^{\prime}({\bf x},{\bf p}) \;=\; \left(\frac{e}{c}\,{\bf A} \;+\; \epsilon\,{\bf p}\right)\bdot\dot{\bf x} \;-\; \left(e\,\Phi \;+\; \epsilon\,\frac{|{\bf p}|^{2}}{2m}\right),
\label{eq:Lag_2}
\end{equation}
which are simply related as $L({\bf x},{\bf p}) \equiv \epsilon^{-1}L^{\prime}({\bf x},{\bf p})$. We note that since the time-dependence of the electromagnetic potentials $(\Phi,{\bf A})$ is not relevant to our discussion, it will, therefore, be ignored in what follows.  In addition, while $\epsilon_{B} \sim \epsilon$, these dimensionless parameters play very different roles, i.e., the particle Lagrangian \eqref{eq:Lag_1} and \eqref{eq:Lag_2} are still meaningful in the case of a uniform magnetic field (where $\epsilon_{B} = 0$.) or time-independent electromagnetic fields. Moreover, the ordering parameter $\epsilon$ is the same dimensionless ordering parameter that appears in the dimensionless equation of motion $\epsilon\,\ov{\bf x}^{\prime\prime} = \ov{\bf x}^{\prime}\btimes\ov{\bf B}$ initially studied by Kruskal \cite{Kruskal_1958,Kruskal_1965}.

As a result of the Lie-transform perturbation analysis, once again based on a definite choice for $\epsilon$ (independent of $\epsilon_{B}$), the guiding-center Lagrangian can also either be expressed according to the charge-renormalization ordering \cite{RGL_1983,Brizard_1989,Cary_Brizard_2009,Tronko_Brizard_2015,Brizard_2023} as
\begin{eqnarray}
L_{\rm gc}({\bf X},p_{\|},\mu,\zeta) &=& \left( \frac{e}{\epsilon c}\,{\bf A} + p_{\|}\,\bhat\right)\bdot\dot{\bf X}  \;-\; H_{\rm gc}({\bf X},p_{\|},\mu) \nonumber \\
 &&+\; \epsilon\,(mc/e)\;\mu \left(\dot{\zeta} - \vb{\cal R}^{*}\bdot\dot{\bf X}\right),
 \label{eq:Lgc_1}
 \end{eqnarray}
where the explicit expression for the guiding-center Hamiltonian $H_{\rm gc}$ is not important in what follows, or, according to the mass-renormalization ordering \cite{Brizard_1995}, as
\begin{eqnarray}
L_{\rm gc}^{\prime}({\bf X},p_{\|},\mu,\zeta) &=& \left( \frac{e}{c}\,{\bf A} + \epsilon\,p_{\|}\,\bhat\right)\bdot\dot{\bf X}  \;-\; \epsilon\,H_{\rm gc}({\bf X},p_{\|},\mu) \nonumber \\
 &&+\; \epsilon^{2}\,(mc/e)\;\mu \left(\dot{\zeta} - \vb{\cal R}^{*}\bdot\dot{\bf X}\right).
 \label{eq:Lgc_2}
 \end{eqnarray}
where $L_{\rm gc}({\bf X},p_{\|},\mu,\zeta)  \equiv \epsilon^{-1}L^{\prime}_{\rm gc}({\bf X},p_{\|},\mu,\zeta)$. In both cases, the guiding-center Lagrangian is independent of the gyroangle $\zeta$ (up to a specified truncated order in $\epsilon$), and, according to Noether's Theorem \cite{Brizard_Lag}, the canonically-conjugate gyroaction $\partial L_{\rm gc}/\partial\dot{\zeta} = \epsilon\,(mc/e)\,\mu$ is a guiding-center invariant (up to that specified truncated order). Once an ordering choice is made (i.e., using either the renormalizations $e/\epsilon$ or $\epsilon\,m$), the $\epsilon$-expansion of the guiding-center Lagrangian has to be consistent with this choice. We note that, with the choice $e = m = c = 1$ used by Littlejohn \cite{RGL_1983}, the guiding-center Lagrangian \eqref{eq:Lgc_1} corresponds exactly to Eq.~(29) obtained by  Littlejohn \cite{RGL_1983} by Lie-transform perturbation method, with the substitution $\Phi \rightarrow \epsilon\,\Phi$

In the guiding-center Lagrangians \eqref{eq:Lgc_1}-\eqref{eq:Lgc_2}, the higher-order correction $-\,(mc/e)\mu\,\vb{\cal R}^{*}\bdot\dot{\bf X}$ involves the vector field  \cite{Tronko_Brizard_2015,Brizard_2023}
\begin{equation}
\vb{\cal R}^{*} \;\equiv\; \vb{\cal R} \;+\; \frac{1}{2}\,\nabla\btimes\bhat,
\label{eq:R_star}
\end{equation}
which includes the gyrogauge vector field $\vb{\cal R} \equiv \nabla\wh{\sf e}_{1}\bdot\wh{\sf e}_{2}$ \cite{RGL_1981,RGL_1983} that is defined in terms of the local fixed unit-vector basis $(\wh{\sf e}_{1},\wh{\sf e}_{2},\bhat \equiv 
\wh{\sf e}_{1}\btimes\wh{\sf e}_{2})$, as well as the term $\frac{1}{2}\,\nabla\btimes\bhat$ that modifies the standard correction $\frac{1}{2}(\bhat\bdot\nabla\btimes\bhat)\,\bhat$ \cite{Brizard_1989} in order to take into account guiding-center polarization \cite{Tronko_Brizard_2015}. While this higher-order correction is absent from Zheng's work \cite{Zheng}, the gyrogauge vector field $\vb{\cal R}$ is needed in the guiding-center Lagrangians \eqref{eq:Lgc_1}-\eqref{eq:Lgc_2} in order to ensure the gyrogauge invariance of the guiding-center equations of motion \cite{RGL_1981,RGL_1983,Brizard_1989} (i.e., the guiding-center Lagrangian dynamics should not only be independent of the gyroangle $\zeta$, but it should also be independent of how the gyroangle is measured in terms of the local perpendicular unit vectors $\wh{\sf e}_{1}$ and $\wh{\sf e}_{2}$). Hence, a proper guiding-center Lagrangian must, at least, include the combination $\dot{\zeta} - \vb{\cal R}\bdot\dot{\bf X}$, which is gyrogauge-invariant \cite{RGL_1983} under the transformation $\zeta \rightarrow \zeta + \psi({\bf X})$, where $\psi({\bf X})$ denotes a locally-defined gyrogauge angle, with $\vb{\cal R} \rightarrow \vb{\cal R} + \nabla\psi$ and $\dot{\zeta} \rightarrow \dot{\zeta} + \dot{\bf X}\bdot\nabla\psi$. The case for time-dependent fields is further discussed in Refs.~\cite{RGL_1981,RGL_1988}, while the importance of the vector field \eqref{eq:R_star} in establishing the faithfulness of the guiding-center representation of particle orbits in nonuniform magnetic fields was recently demonstrated for the case of axisymmetric magnetic geometries \cite{Brizard_2023}.
 
 In his critique of standard Lie-transform perturbation analysis, and without explicitly displaying the dimensionless ordering parameter $\epsilon$ upon which it is to be based, Zheng \cite{Zheng} mistakenly proceeds to compare the guiding-center Lagrangians \eqref{eq:Lgc_1} and \eqref{eq:Lgc_2}, derived with different renormalization orderings, and concludes that, when the guiding-center Lagrangian \eqref{eq:Lgc_2} is truncated at first order, the term $\epsilon^{2}(mc/e)\mu\,\dot{\zeta}$ disappears, while the term $\epsilon\,(mc/e)\mu\,\dot{\zeta}$ remains in the guiding-center Lagrangian \eqref{eq:Lgc_1}, although it is still a second-order term compared to the lowest order term appearing at $\epsilon^{-1}$. However, Zheng seems to be unaware that the guiding-center Lagrangian \eqref{eq:Lgc_1}, which was derived without Lie-transform perturbation method by Cary and Brizard \cite{Cary_Brizard_2009} in what Zheng calls the {\it direct} method, was also derived by Lie-transform perturbation method by Littlejohn \cite{RGL_1983}, Brizard \cite{Brizard_1989}, and Tronko and Brizard \cite{Tronko_Brizard_2015}. 
 
 More importantly, Zheng argues that, in contrast to the $\epsilon$-ordering scalings displayed in the guiding-center Lagrangians \eqref{eq:Lgc_1}-\eqref{eq:Lgc_2}, the terms $p_{\|}\bhat\bdot\dot{\bf X}$ and $(mc/e)\;\mu\,\dot{\zeta}$ must appear at the same order in a modified guiding-center perturbation expansion, which leads him to construct a completely unnecessary (and nonsensical) modification of Lie-transform perturbation theory. However, this modified ordering is clearly inconsistent with the property of gyrogauge invariance based on the following argument. First, by momentarily hiding the $\epsilon$-ordering scalings in Eqs.~\eqref{eq:Lgc_1}-\eqref{eq:Lgc_2}, the guiding-center Lagrangian can be written as
 \begin{eqnarray}
 L_{\rm gc}({\bf X},p_{\|},\mu,\zeta) &=& \left[ \frac{e}{c}\,{\bf A} \;+\; p_{\|}\,\bhat \;-\; (mc/e)\;\mu\,\vb{\cal R}^{*} \right]\bdot\dot{\bf X} \nonumber \\ 
  &&+\; (mc/e)\;\mu\,\dot{\zeta} \;-\; H_{\rm gc}({\bf X},p_{\|},\mu),
  \label{eq:Lgc_0}
  \end{eqnarray}
  where we have combined the gyrogauge-correction term $-\; (mc/e)\;\mu\,\vb{\cal R}^{*}$, omitted in Zheng's work \cite{Zheng}, with the spatial components $(e/c){\bf A} + p_{\|}\bhat$. Here, we clearly see that these spatial components satisfy the following ordering $\epsilon^{-1} \gg 1  \gg \epsilon$ \cite{footnote}. Hence, after restoring the $\epsilon$-ordering scalings of the spatial components in Eq.~\eqref{eq:Lgc_0}, we obtain
  \begin{eqnarray}
 L_{\rm gc}({\bf X},p_{\|},\mu,\zeta) &=& \left[ \frac{e}{\epsilon c}\,{\bf A} \;+\; p_{\|}\,\bhat \;-\; \epsilon\,(mc/e)\;\mu\,\vb{\cal R}^{*} \right]\bdot\dot{\bf X} \nonumber \\ 
  &&+\; \delta\,(mc/e)\;\mu\,\dot{\zeta} \;-\; H_{\rm gc}({\bf X},p_{\|},\mu),
  \label{eq:Lgc_a}
  \end{eqnarray}
  where we have also introduced a dimensionless ordering parameter $\delta$ for the gyromotion term $(mc/e)\;\mu\,\dot{\zeta}$. Next, we note that the guiding-center Lagrangian \eqref{eq:Lgc_a} now contains the gyrogauge combination
  \[ (mc/e)\;\mu \left( \delta\,\dot{\zeta} \;-\frac{}{} \epsilon\,\vb{\cal R}\bdot\dot{\bf X}\right), \]
 which is gyrogauge invariant only if $\delta = \epsilon$ (and not $\delta = 1$ as proposed by Zheng \cite{Zheng}), i.e., the term $(mc/e)\;\mu\,\dot{\zeta}$ must appear at one order higher than $p_{\|}\bhat\bdot\dot{\bf X}$ in a perturbation expansion leading to a gyrogauge-invariant guiding-center Lagrangian theory, based on either Eq.~\eqref{eq:Lgc_1} or Eq.~\eqref{eq:Lgc_2}. The ordering $\delta = \epsilon$ in Eq.~\eqref{eq:Lgc_a} is, therefore, entirely consistent with the renormalization $m/e \rightarrow \epsilon\,m/e$ of the mass-to-charge ratio used (either implicitly \cite{RGL_1983,Brizard_1989} or explicitly \cite{Cary_Brizard_2009,Tronko_Brizard_2015}) in previous works as the consistent basis for applications of the standard Lie-transform perturbation analysis.
 
 In conclusion, the standard Lie-transform perturbation method \cite{RGL_1982} does not need to be modified in its applications to guiding-center theory \cite{RGL_1983,Cary_Brizard_2009} and, fortunately, the modified Lie-transform perturbation method proposed by Zheng \cite{Zheng}  will not be needed in deriving a modified nonlinear gyrokinetic theory \cite{Brizard_Hahm_2007}. The paper by Zheng \cite{Zheng} reminds us that perturbation theory relies on a well-defined dimensionless ordering parameter 
 $\epsilon \ll 1$, followed by a rigorous algorithm (e.g., Lie-transform perturbation theory) that allows terms to be computed at arbitrary order.

\vspace*{0.2in}

\acknowledgments

The present work was supported by the National Science Foundation grant PHY-2206302.

\vspace*{0.1in}

\noindent
{\bf AUTHOR DECLARATIONS}

\vspace*{0.1in}

\noindent
{\bf Conflict of Interest} 

\vspace*{0.1in}

The author has no conflicts to disclose.

\end{document}